\documentclass[twocolumn,showpacs,amsmath,amssymb,prl,a4paper]{revtex4}
\usepackage{graphicx}% Include figure files
\usepackage{dcolumn}% Align table columns on decimal point
\usepackage{bm}% bold math

\newcommand{\TT}{\mathcal{T}}
\newcommand{\TTn}{\mathcal{T}}

\newcommand{\hh}{\mathfrak{h}}
\newcommand{\ff}{f}

\newcommand{\rme}{\mathrm{e}}
\newcommand{\re}{\mathop\mathrm{Re}}

\begin{document}

\title{Critical Behavior of Light}
\author{Rafi Weill, Amir Rosen, Ariel Gordon, Omri Gat and Baruch Fischer}
% \altaffiliation[Also at ]{Physics Department, XYZ University.}
%\affiliation{Department of Electrical Engineering, Technion, Haifa 32000, Israel }
\address{Department of Electrical Engineering, Technion, Haifa
32000, Israel }

\date{\today}
\begin{abstract}

Light is shown to exhibit critical and tricritical behavior in
passive mode-locked lasers with externally injected pulses. It is
a first and unique example of critical phenomena in a
one-dimensional many body light-mode system. The phase diagrams
consist of regimes with continuous wave, driven para-pulses,
spontaneous pulses via mode condensation, and heterogeneous
pulses, separated by phase transition lines which terminate with
critical or tricritical points. Enhanced nongaussian fluctuations
and collective dynamics are observed at the critical and
tricritical points, showing a mode system analog of the critical
opalescence phenomenon. The critical exponents are calculated and
shown to comply with the mean field theory, which is rigorous in
the light system.

\end{abstract}

\maketitle

%\maketitle

%\nosections

Statistical light-mode dynamics (SLD) \cite{GFPRL,GGF} that is
based on statistical mechanics, provides a powerful approach for
the study of complex nonlinear light systems, while serving as a
new statistical mechanics paradigm. It was developed to treat and
solve long standing questions in laser physics. A prime example is
the laser pulsation threshold, a phenomenon in passive mode
locking that attracted much attention \cite{self-starting}.
Passive mode-locking occurs when a saturable absorber element is
placed in the cavity, driving the laser to generate extremely
short light pulses. The transition mechanism from a continuous
wave (cw) to pulsation of such lasers was shown via SLD
\cite{GFPRL,GGF} to be an intrinsic property of the many
interacting mode system, ruled by the balance between the
nonlinear interaction induced by the saturable absorber and the
randomizing effect of noise. The theory, verified by experimental
study, demonstrated \cite{steps} the existence of first order
phase transitions between disordered (cw) and ordered (locked)
mode phases, as the noise (``temperature") or the laser power were
varied. Thus noise alone stabilizes the cw state, showing a noise
induced phase transition. \cite{jordi}.

In the present Letter we report on findings of \emph{critical} and
\emph{tricritical} phenomena in the many light-mode system. It is
a first example of such behavior with light, that also provides a
physical realization of a strict one-dimensional many body system.
For criticality to appear in the laser mode system we add to it an
external driving field which is a counterpart of the external
magnetic field in magnets and the pressure in gas-liquid-solid
systems \cite{stanley}. It is achieved by injecting the laser with
pulses from an external source, which in the simple case matches
the repetition rate of the laser. When the injection is weak, the
ordering phase transition persists, shifted to higher
``temperature'', with a``para-pulses" phase (pulses induced and
driven by the external injection). However, the phase transition
line terminates in a \emph{critical point}, where the distinction
between para-pulses and spontaneous pulses disappears, similarly
to the vapor-liquid critical point. Thus, by increasing the
external injection, it is possible to obtain mode locking
{smoothly} from cw. Near the critical point the system exhibits
the familiar critical phenomena, including divergence of response
coefficients, characterized by universal critical exponents, and
nongaussian critical fluctuations, enhanced by a factor of
$N^{1/4}$, where $N$ is the number of active modes, compared to
normal fluctuations. The latter could mean a two orders of
magnitude fluctuations enhancement in practical systems. It is a
light mode system analog of the critical opalescence phenomenon.

When the injection repetition rate is higher than that of the laser,
the high- and low-temperature phases are characterized by equal and
unequal pulse powers, respectively, which cannot be smoothly connected.
However, beyond a threshold injection level the transition
becomes \emph{continuous} rather than first order. The two phase
transition lines meet at a \emph{tricritical} point, often found in
systems with interaction competing with external driving \cite{tricritical},
around which tricritical behavior is observed, with its distinct set of
universal exponenets, and the tricritical fluctuations are enhanced by a factor of $N^{1/3}$.

The inclusion of external injection has other important
consequences. For example, as in the vapor-liquid case, field
induced ``condensation", that is in our case mode ordering and
pulsation, can be sustained to some extent when the field is
removed, only below the critical point. Another practical aspect
of external injection is the possibility to overcome basic
deficiencies in passively mode locked lasers, their low repetition
rate and timing jitter \cite{margalit1}. Our theory provides a
basic and quantitative understanding of the way that an external
field locks the laser pulsation, and the threshold power needed
(or maximum noise allowed) for various operation scenarios to
happen.

We perform our theoretical study in the framework of the
coarse-grained model of SLD, representing the cavity electric
field envelope $\psi$ by a single variable in an interval whose
length is of the order of the pulse width. The derivation of the
model from the passive mode locking master equation
\cite{HausReview} has been discussed before
 \cite{GFPRL,GGF,steps}, and it shows good quantitative agreement with theoretical
 studies of the master equation as well as with experiments.

The dynamics of the the field variable in interval number $n$ is expressible as
\begin{equation}\label{eq:psi}
\partial_t\psi_m=-\frac{\partial H}{\partial \psi_m^*}+g\psi_m+
\eta_m(t)\ .
\end{equation}
The ``Hamiltonian" $H$ with injection strength $h$ takes the form
\begin{equation}
\label{eq:hamiltonian_time}
H=-N\re\sum\limits_{m=1}^N(\frac{\gamma}{2}|\psi_m|^4+2h_m\psi_m^*)\
,
\end{equation}
where $N\gg1$ is the number of active modes, $\gamma$ is the
coefficient of saturable absorption, and $h_n$ is the external
injection at site $n$. $h_n$ will be assumed to take nonzero
values only at a small number $n$ of intervals. $g$ is the overall
net gain, which can be assumed without loss of generality
\cite{GGF} to set the intracavity power
$\|\psi\|^2=\sum_n|\psi_n|^2$ to a fixed value $P$, in which case
$g$ becomes a Lagrange multiplier for the constraint. The random
term $\eta$, representing noise from spontaneous emission and
other sources, is modelled by a (complex) Gaussian white noise
with covariance
$\left<\eta_n^*(t)\eta_m(t')\right>=2T\delta_{nm}\delta(t-t')$.

The invariant measure of Eq. (\ref{eq:hamiltonian_time}) is a Gibbs equilibrium
distribution \cite{Risken,GFPRL}
\begin{equation}\label{eq:inv}
\rho[\psi]=\frac{1}{Z}\rme^{-H[\psi]/T}\delta(\|\psi\|^2-P)\ ,
\end{equation}
with the partition function
\begin{equation}\label{eq:z}
Z=\int [d\psi][d\psi^*] \rme^{-H[\psi]/T}\delta(\|\psi\|^2-P)\ .
\end{equation}

As in previously studied cases of SLD \cite{GFPRL,GGF,steps}, when
$N\gg1$ the  invariant measure is concentrated on configurations
where all but a finite number of the $\psi$ variables are
$O(N^{-1/2})$. Here the intervals where $\psi=O(1)$ are precisely
the $n$ intervals which are subject to external injection. The
free energy $F=-T\log Z$ is then given by
$F=N\min_{\psi_1,\ldots,\psi_n}f_n(\psi_1,\dots,\psi_n)$, where
\begin{equation}\label{eq:fmin}
f_n=-\sum_{m=1}^n (\frac{\gamma}{2}|\psi_m|^4+2\re
h_m\psi_m^*)+T\log(P-\sum_{m=1}^n|\psi_m|^2) \ .
\end{equation}
We establish Eq. (\ref{eq:fmin}) using the results of \cite{GGF}
for  the partition function $Z_0$ for the $h=0$ case,
\begin{equation}\label{eq:f0min}
Z_0\sim\int d^2\psi\rme^{[N\gamma|\psi|^4+T\log(P-|\psi|^2)]/T}
\end{equation}
asymptotically for large $N$. We can proceed immediately to
perform the integration in Eq. (\ref{eq:z})  over the $N-n$
intervals where $h=0$
\begin{eqnarray}\label{eq:zm}
&&Z\sim\int d^2\psi\prod_m d^2\psi_m\\\nonumber
&&\quad\times\rme^{-\frac{1}{T}(H(\psi_1,
\ldots,\psi_n)+N\gamma|\psi|^4+NT\log(P-\sum_{m=1}^n|\psi_m|^2-|\psi|^2)}
\end{eqnarray}
The exponent in the integrand in Eq.\ (\ref{eq:zm}) is
proportional to the large  parameter $N$, and therefore the
integration is concentrated near the global minimum of the
integrand. It is straightforward to verify that the minimum is
always obtained when $\psi=0$, which then implies that $F=N\min
f_n$.

Moments of the pulse strength can be obtained in the standard
manner by taking  derivatives of the free energy. Alternatively we
may obtain directly the probability distribution function of the
pulse amplitudes by integrating out the $N-n$ unforced variables
in Eq. (\ref{eq:inv}) getting
\begin{equation}\label{eq:pofpsi}
{\sf P}(\psi_1,\ldots,\psi_n)\sim\rme^{Nf_n(\psi_1,\ldots,\psi_n)}
\ .\end{equation} The width of the distribution tends to zero in
the thermodynamic limit $N\to\infty$,  which shows that the pulse
amplitudes are thermodynamic observables.
% The distribution Eq. (\ref{eq:pofpsi}) is discussed further below in the context of critical fluctuations.

We henceforth specialize to the case that all nonzero injection
values are of  the same magnitude $h$, that is appropriate for
injection of pulses from a source with a repetition rate $n$ times
faster than that of the laser. For the minimization problem there
is no loss of generality in assuming that the injection values are
all real, since the minimizing $\psi$ values have the same phase
as the corresponding $h$ values. By a simple rescaling of the
variables the free energy $f_n$ can be reduced to
\begin{equation}\label{eq:f_time}
 \ff=-\sum\limits_{m=1}^n(\frac 1 2 x_m^4+\hh x_m)-\TT\log{\left(1-\sum\limits_{m=1}^nx_m^2\right)}\ ,
\end{equation}
where $x_m$'s are real and positive. It follows from Eq.\
(\ref{eq:f_time})  that thermodynamics depends on only two
dimensionless parameters, the reduced temperature
$\TT=\frac{T}{\gamma P^2}$ (the inverse of of the interaction
strength \cite{GFPRL}) and reduced driving $\hh=\frac{2h}{\gamma
P^{3/2}}$. $\bar x_m$, the minimizers of $\ff$, are related to the
expectation values (in the invariant measure) of the pulse powers
by $\bar x_m^2=\left<|\psi_m|^2\right>/P$.

The study of thermodynamics and critical behavior is now reduced
to the analysis of the function $\ff$ and its minima. We note that
for any values of the parameters, minima can occur only at
configurations where at least $n-1$ of the $\bar x_m$'s are equal,
with value $\overline y$, and the other minimizer, which we denote
by $\overline x$ is greater than or equal to $\overline y$.
Accordingly the free energy is obtained from
\begin{eqnarray}
 \ff(x,y)=-\frac 1 2
 \left[x^4-(n-1)y^4\right]-\hh[x+(n-1)y]\qquad&&\nonumber\\-\TT\log[{1-x^2-(n-1)y^2}]
\label{eq:f_simple}\end{eqnarray}

It is instructive to consider the above results in the
$\mathbf{k}$ or mode space $a_k$, the discrete  Fourier transform
of $\psi_n$. The Hamiltonian reads
\begin{equation}\label{eq:hamiltonian_freq}
      H=-\frac{\gamma}{2}\sum\limits_{j-k+l-m=0}a_j a_k^* a_l
             a_m^*-2\re\sum_k\tilde h_k a_k^*\ ,
\end{equation}
where $\tilde h_k$ takes nonzero value for $k$ being integer
multiples of $n$. Therefore, when $n>1$, the set of modes consists
of two types, with and without the presence of the external field.
The hamiltonian in Eq.~(\ref{eq:hamiltonian_freq}) is analogous to
the one of an antiferromagnet placed in an external homogeneous
magnetic field \cite{metamagnets}, but with the roles of the
interaction term and the driving term reversed. Namely, the
driving acts on a subset of the modes, but the interaction tends
to align all modes in the same amplitude and phase. The
consequences of this competition are derived below. The free
energy can be expressed in the mode representation using the mode
amplitude expectation values $\left<a_f\right>$ and
$\left<a_u\right>$ of those with (forced) and without electric
field (unforced), respectively. The relations are
$x=\frac1n(\left<a_f\right>+(n-1)\left<a_u\right>)$ and $y=\frac 1
n
  (\left<a_f\right>-1\left<a_u\right>)$ and the free energy is
\begin{eqnarray}
 &&\ff=-\frac{1}{2n^3}[\left<a_f\right>^4+(n-1)(n^2-3n+3)\left<a_u\right>^4\nonumber\\&&\qquad
 +6(n-1)\left<a_f\right>^2\left<a_u\right>^2+4(n-1)(n-2)\left<a_f\right>\left<a_u\right>^3]
 \nonumber\\&&\qquad-\hh \left<a_f\right>-\TT\log{(1-\frac 1 n \left<a_f\right>^2
-\frac{n-1}{n}\left<a_u\right>^2)} \label{eq:f}\end{eqnarray}

\begin{figure}[htb]
\centering
\hbox{\hskip1cm\includegraphics[width=6.5cm,height=4cm]{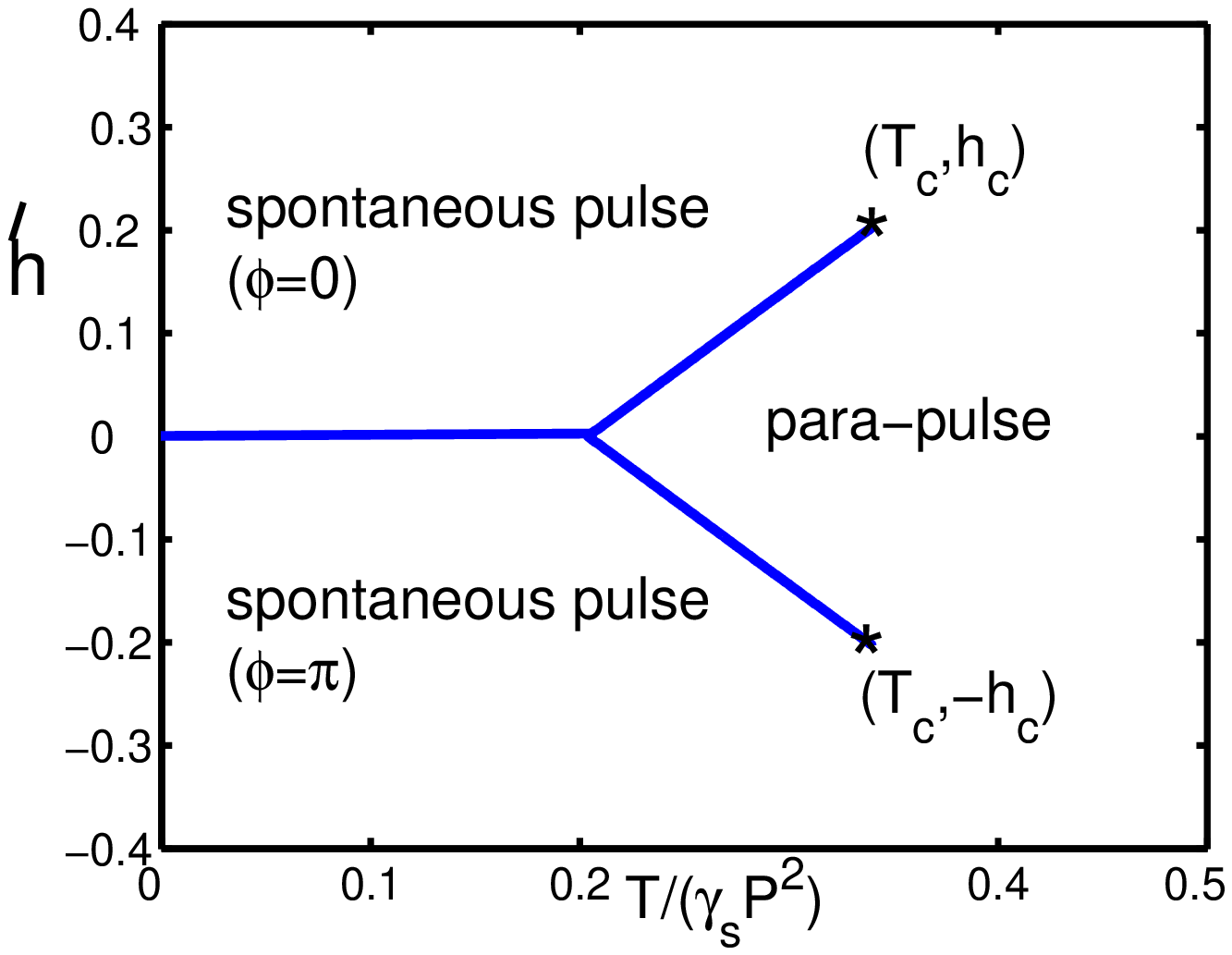}}
\centering
\hbox{\hskip1cm\includegraphics[width=6.5cm,height=4cm]{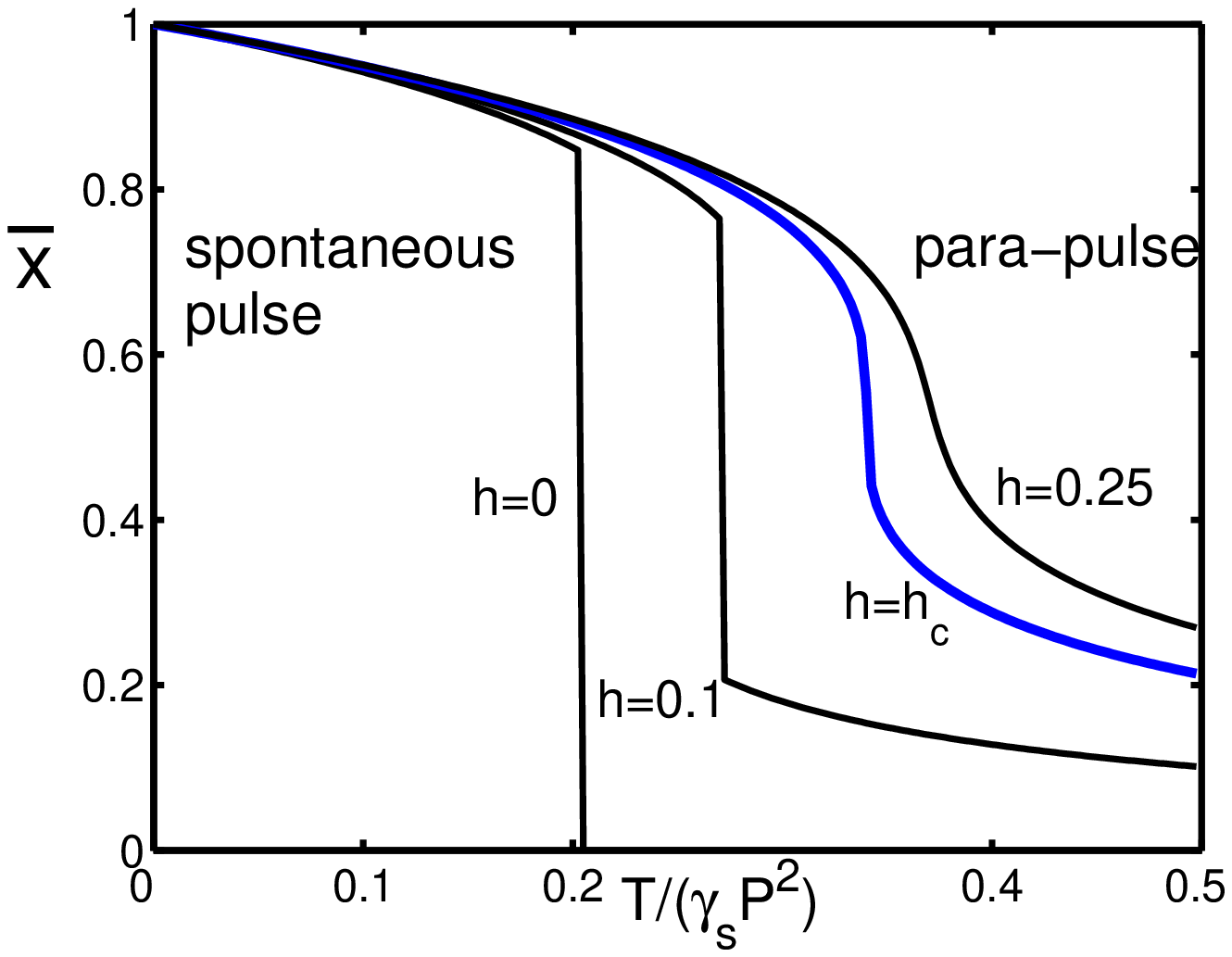}}
\caption{Upper figure: The phase diagram n=1 (homogeneous
injection to all modes that matches the cavity repetition rates).
The lines gives the first order phase transition curve, that
terminate at the critical points. The optical phase $\phi$ follows
the phase of $\hh$. Lower figure: Pulse power vs. $\TT$ for
different values of external injection $\hh$.
\label{fig:external1_phase}}
\end{figure}

For the analysis below we use the real space formulation. We
consider first the case of $n=1$, where there is a single pulse
(the external field is applied on all modes), and $f$ (Eq. 10)
depends on the single variable $x$. For zero and small values of
$\hh$, $\ff$ has a minimum $x_1$ near zero and, for small enough
$\TT$ another minimum $x_2>x_1$ below 1. For such $\hh$ there is a
threshold temperature $\TT_1(\hh)$ where $\ff(x_1)=\ff(x_2)$. As
$\TT$ is decreased through this line $\bar x$ jumps from $x_1$ to
$x_2$ in a first order phase transition. When $\hh>0$ the jump is
between two pulsed states, but  the high-temperature phase pulses
are driven pulses whose power decreases smoothly to zero when
$\hh\to0$. We term this phase ``para-pulse" because of its
resemblance with the paramagnetic phase of magnets above the Curie
point. For sufficiently strong $\hh$, on the other hand, $\ff$
always has a single  minimum $\bar x$ between zero and one, which
decreases smoothly from one to zero as $\TT$ is increased. Thus,
the coexistence line $\TT_1(\hh)$ terminates at a critical point
$(\hh_c,\TT_c)$. The phase diagram is shown in the upper part of
Fig. \ref{fig:external1_phase}.

The lower part of Fig.~\ref{fig:external1_phase} shows $\bar x$ as
a function of $\TT$ for several  $\hh$ values. $\bar x$ undergoes
a jump for $\hh<\hh_c$, and displays an infinite slope at the
critical point---a manifestation of the critical divergence of the
susceptibility. The critical point itself is characterized by the
vanishing of the first three derivatives of $\ff$, which gives
three polynomial equations for the three unknowns $\bar x_c$,
$\hh_c$ and $\TT_c$. The equations can be solved explicitly by
radicals giving $\bar{x}_c\approx0.53$, $\hh_c\approx0.2035$,
$\TT_c\approx0.34$.

One may define the usual critical exponents \cite{stanley}
 $\beta$, $\gamma,$ $\delta$ by
$(\bar{x}-\bar{x}_c)|_\textrm{coexistence}\sim(\TT_c-\TT)^\beta$,
$\chi={\partial \overline{x}}/{\partial
\hh}\sim|\TT-\TT_c|^{-\gamma}$, and
$(\overline{x}-\overline{x}_c)|_{\TTn=\TTn_c}\sim(\hh-\hh_c)^{1/\delta}
$. The exponents, as well as the nonuniversal amplitudes can be
calculated by the standard procedure of expanding $\ff$ up to
third order near the critical point \cite{stanley}, which yields
the classical mean field exponents $\beta=1/2$, $\gamma=1$,
$\delta=3$, expectedly, since mean field theory applies to our
system.

The fluctuation-dissipation relations naturally hold in the laser
mode system; it follows from Eq. (\ref{eq:z}) that
\begin{equation}
\mathrm{Var}(\psi)=\frac2N\frac{\partial\left<\psi\right>}{\partial h}
=\frac2N\chi\ ,
\end{equation}
that is, the critical exponent $\gamma$ also describes the
divergence  of pulse power fluctuations near the critical point.
To study the fluctuations \emph{at} the critical point we turn
back to Eq. (\ref{eq:pofpsi}), which in the present context is a
probability distribution for the single pulse amplitude. Letting
$x=P^{-1/2}|\psi|$, the criticality condition implies that for
$T=T_c$, $h=h_c$
\begin{equation}
{\sf P}(x)\sim\rme^{N(a(x-x_c)^4+O(x-x_c)^5)}\ ,
\end{equation}
where $a$ is $O(1)$. The fluctuations distribution is nongaussian,
and the \emph{scale} of the critical fluctuations is
$O(N^{-1/4})$, stronger by a factor of $N^{1/4}$ than normal
fluctuations. The critical fluctuations are much larger than the
typical amplitude of the continuum background, $O(N^{-1/2})$. It
follows that the fluctuations of the continuum background are
\emph{correlated}, being the SLD analog of the critical
\emph{opalescence} phenomenon.

The thermodynamics with $n>1$ is qualitatively different. The
external injection encourages the formation of $n$ equal pulses,
clashing with the tendency of the saturable absorber to form a
single strong pulse. As a result, the phase diagram consists of
unequal pulse phase for weak noise and weak injection, and an
equal pulse phase for strong noise or strong injection. As the two
phases are characterized by different symmetries, there can be no
smooth transition between them, and they are separated by a phase
transition line. However, the phase transition may be continuous
or first order, depending on whether the transition order
parameter $q=(x-y)/\sqrt{x^2+y^2}$ is continuous or jumps to a
nonzero value at the transition.

We consider in detail the case $n=2$ where both behaviors occur.
It is more convenient to express $\ff$ of Eq.\ (\ref{eq:f_simple})
in terms of $q$ and $p=x^2+y^2$
\begin{equation}\label{eq:fpq}
\ff(p,q)=-\frac{p^2}{2}(1+2q^2-q^4)-
\hh\sqrt{2p\Big(1-\frac{q^2}{2}\Big)}-
\TT\log(1-p)\
\end{equation}
to be minimized over $p$ and $q$. The values of the minimizers
$\bar{p}$ and $\bar{q}$, giving their "thermal" averages vs.
$\TT$, are given in Fig. 2. $\ff$ is manifestly symmetric in $q$,
since the pulse amplitudes $x$ and $y$ play symmetric roles,
wherefrom it follows that $\ff$ is always stationary with respect
to $q$ when $q=0$. The condition $\partial_q\ff=0$ has another
solution
%\begin{equation}\label{eq:pq}
$p^{3}=\frac{\hh^2}{4}(q^2-1)^2(2-q^2),$
%\end{equation}
and the global minimum of $\ff$ is reached in one of these
configurations, the first corresponding to equal and the second to
unequal pulse mode locking.

Straightforward analysis shows that for large $\TT$ the function
$\ff$ has a single minimum, which occurs at $q=0$. For large $\hh$
this situation persists as $\TT$ is lowered until at
$\TT=\TT_b\equiv\frac{3}{4}\hh^{2/3}(2-\hh^{2/3})$ the minimum
becomes a saddle and two minima with nonzero $q$ form, i.e., $q$
undergoes a continuous phase transition, see Fig.\
\ref{fig:phase}. For small $\hh$, on the other hand, nonzero $q$
minima appear for $\TT>\TT_b(\hh)$, and at $\TT_1(\hh)$ exchange
stability with the $q=0$ minimum in a first order phase
transition, also shown in Fig.\ \ref{fig:phase}.

\begin{figure}[tb]
\hbox{\includegraphics[width=4.25cm]{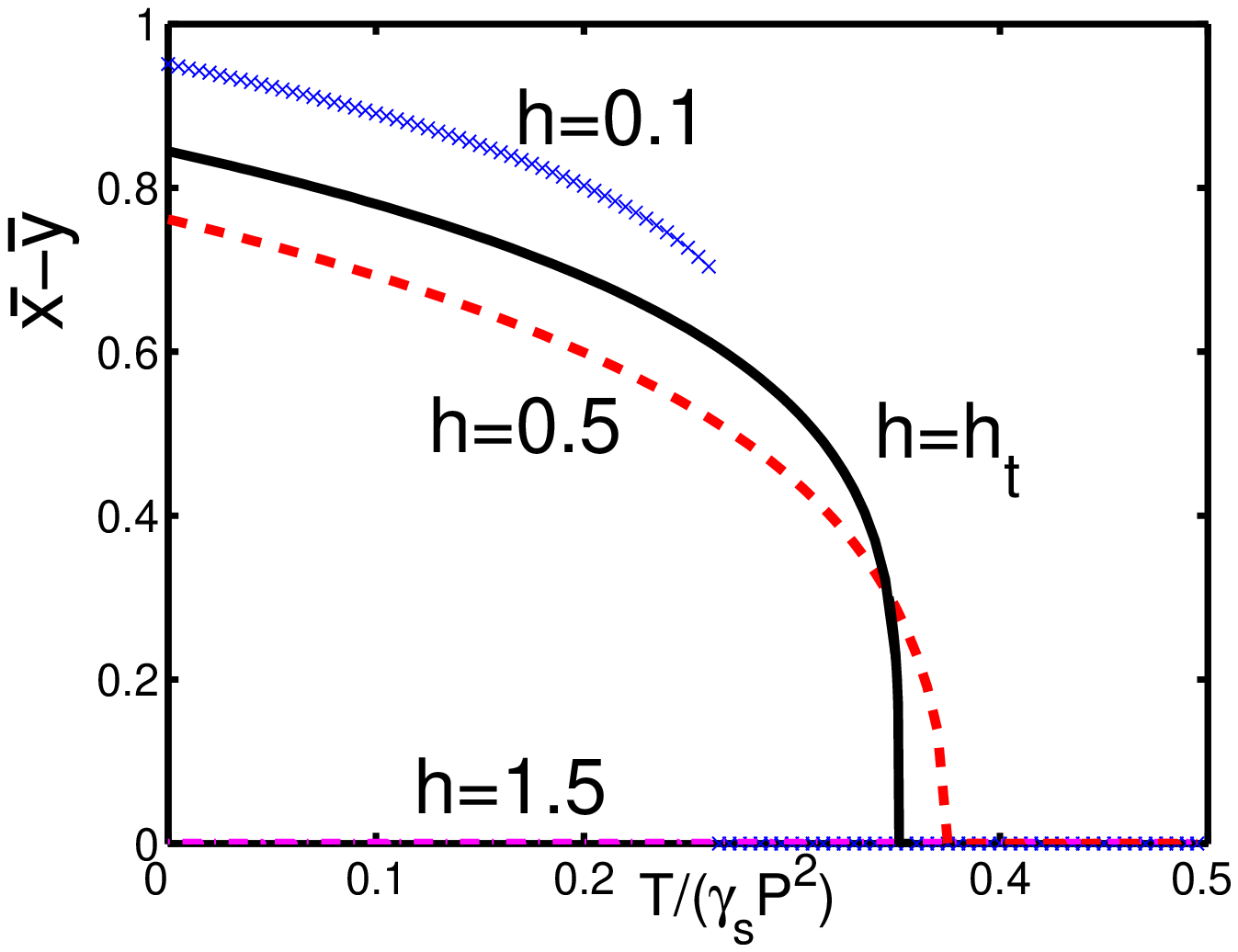}
\hskip0.1cm\includegraphics[width=4.25cm]{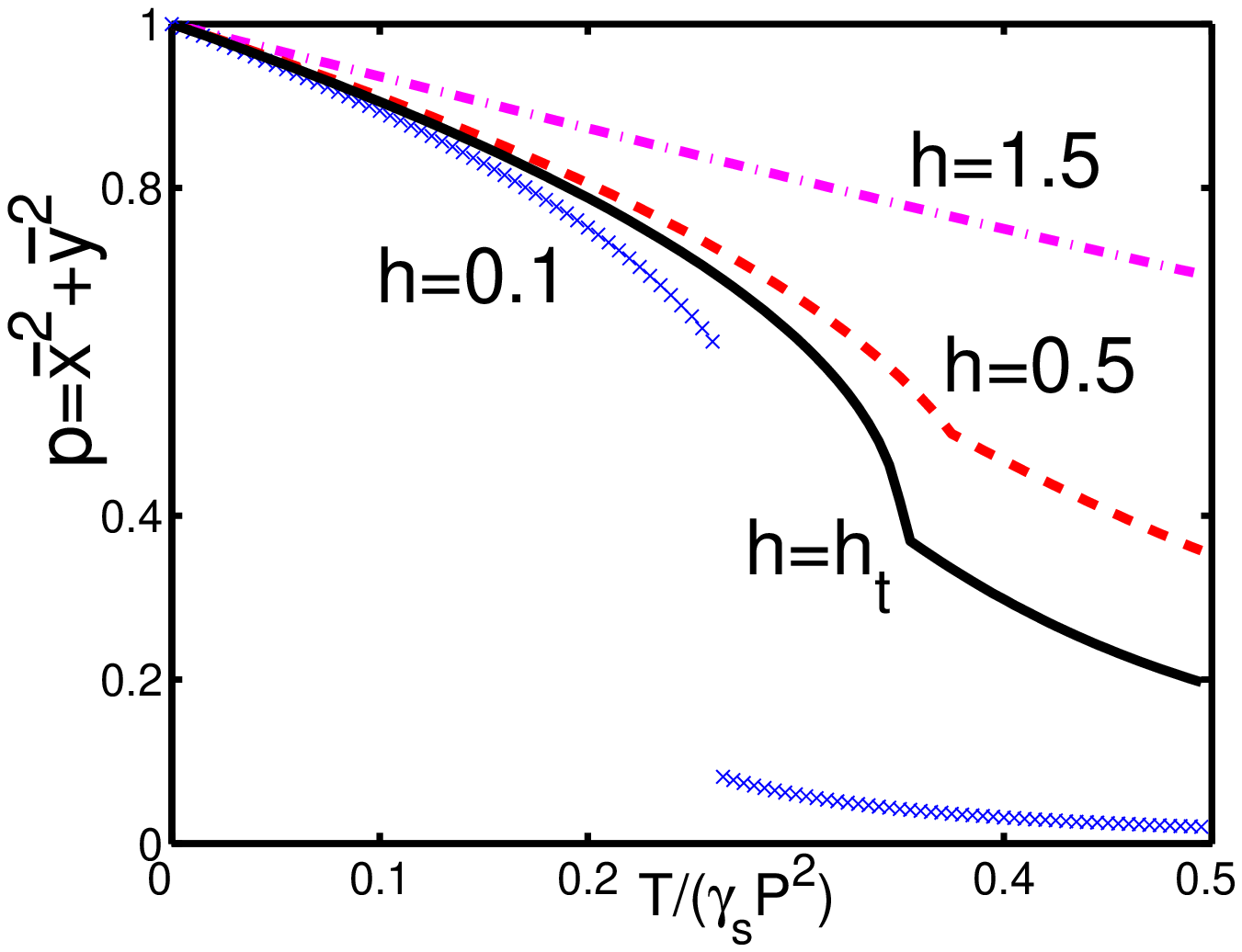}}
\caption{$(\overline{x}-\overline{y})$ (left) and
$\overline{x}^2+\overline{y}^2$ (right) vs. the normalized
temperature for different values of $\hh$ for injection repetition
rate twice the laser repetition rate. The nature of the phase
transition changes at $\hh=\hh_t$} \label{fig:external2}
\end{figure}

\begin{figure}[htb]
\hbox{\includegraphics[width=4.25cm]{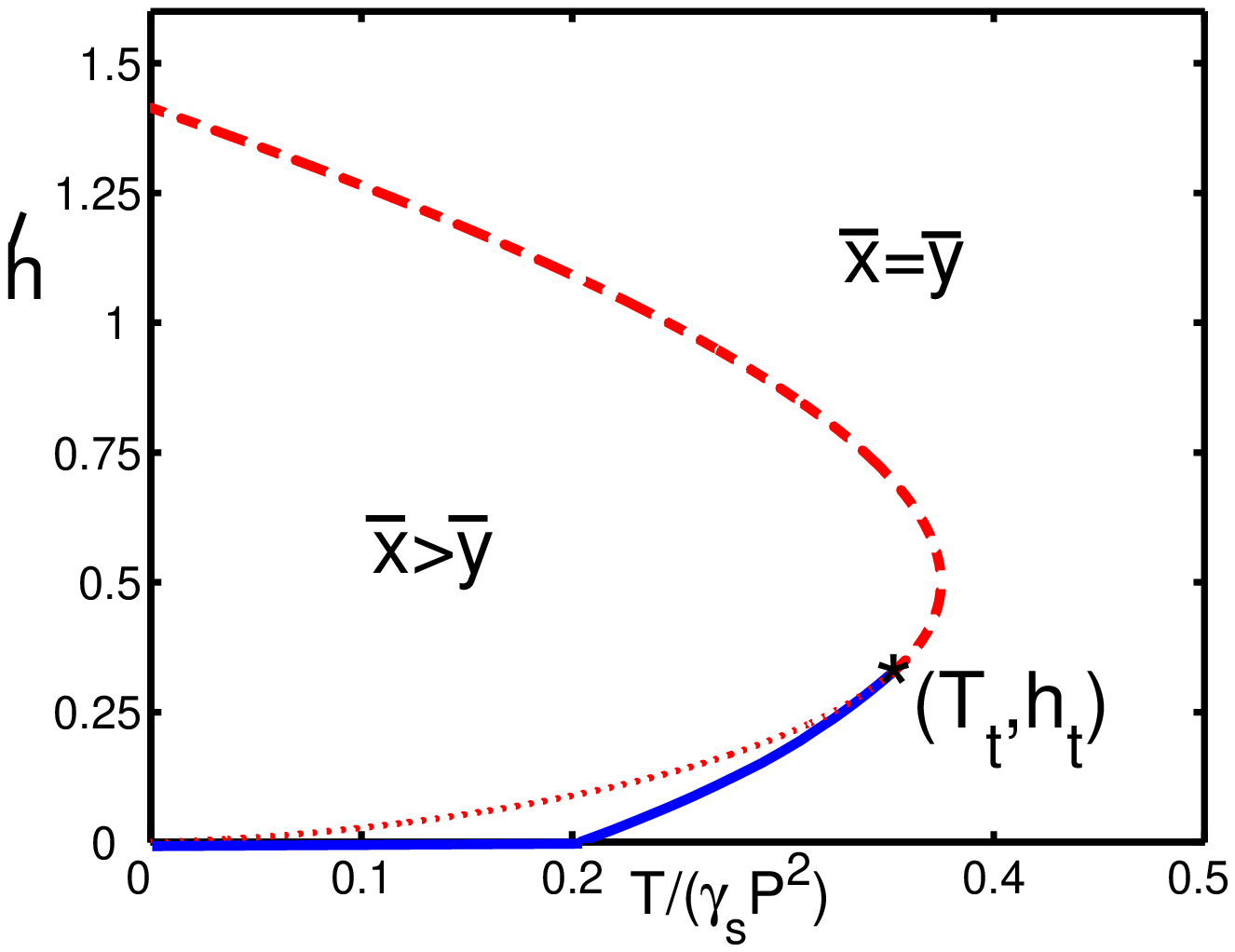}
\hskip0.1cm\includegraphics[width=4.25cm]{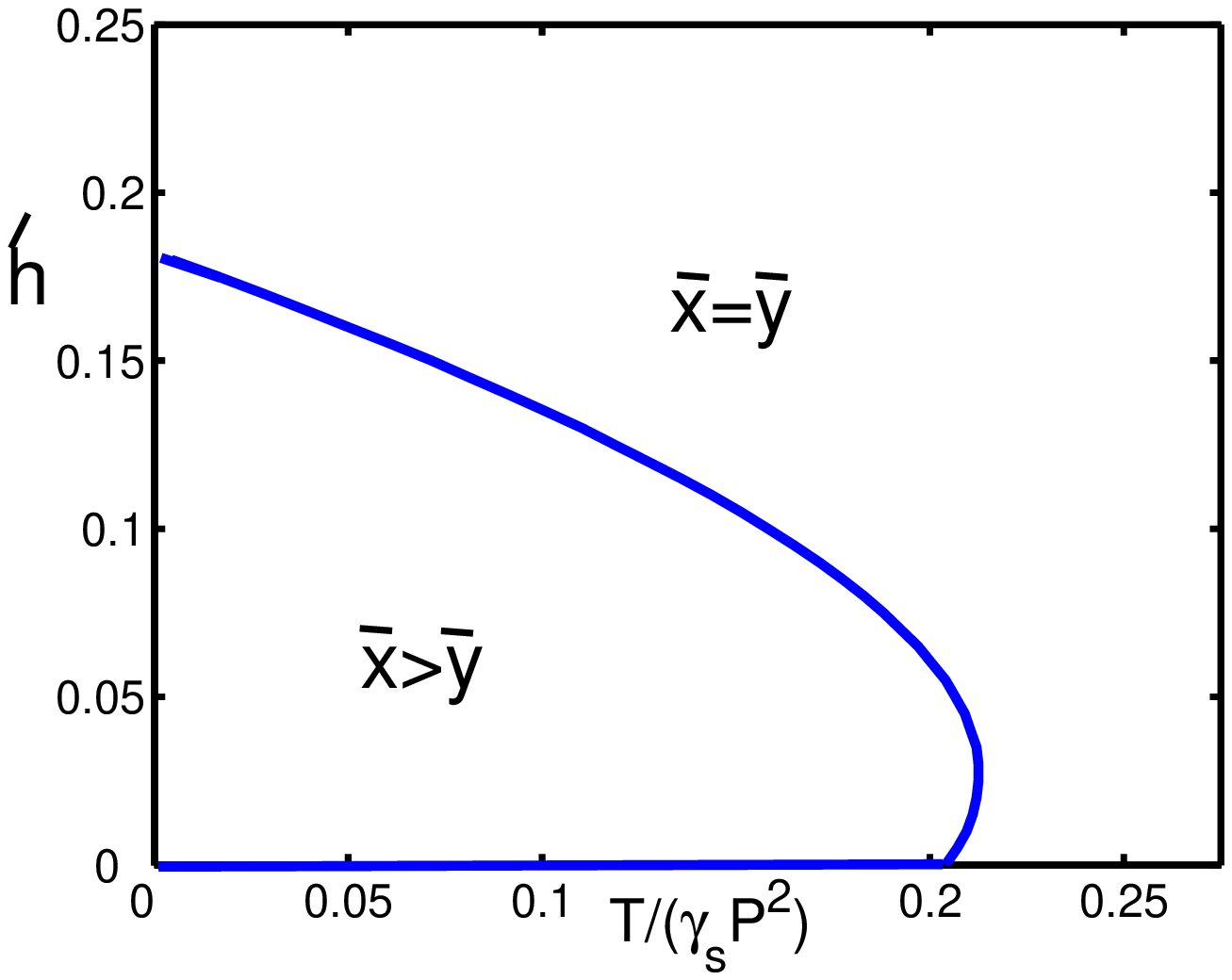}}
\caption{The $\hh-\TT$ phase diagrams for repetition rate ratios
$n=2$ (left) and $n=20$ (right). The bold line is a first order
phase transition curve. The dotted and dashed lines form together
the bifurcation curve $\TT_b(\hh)$, of which the first is a line
of continuous phase transition.\label{fig:phase}}
\end{figure}

The intersection of the line of first order phase transition
$\TT_1(\hh)$ and the line of continuous phase transition
$\TT_b(\hh)$ can be shown to occur at
$(\TT_t,\hh_t)=(\frac{3^{3/2}}{16},\frac{45}{128})$.
$(\TT_t,\hh_t)$ is a \emph{tricritical} point \cite{tricritical},
with symmetric tricritical phenomena, and the phase diagram Fig.\
\ref{fig:phase} is quite similar to that of metamagnets
\cite{metamagnets}, where tricritical behavior is known to occur.
In particular we may define tricritical exponents such as
$\beta_t$ and $\beta_{2t}$ associated with nonsymmetric  (e.g.
$q$) and symmetric (e.g. $p$) fields respectively by
$q\sim(\TT_t-\TT)^{\beta_t}$ and
$p-p_t\sim(\TT_t-\TT)^{\beta_{2t}}$ near the tricritical point,
see Fig. \ref{fig:external2}. As before, the exponents take the
classical values $\beta_t=1/4$, $\beta_{2t}=1/2$
\cite{tricritical}. Near the continuous phase transition line
there are ordinary critical phenomena, for example
$q\sim(\TT_b-\TT)^\beta$ and $p-p_b\sim(\TT_b-\TT)^{\beta_2}$,
where $\beta=1/2$ and $\beta_2=1$. At the tricritical point
fluctuations are enhanced by a factor $N^{1/3}$ compared to normal
fluctuations.

When the ratio of the repetition rates $n$ is three or larger one
can show that the transition between the equal and unequal pulse
phases is always first order; a typical phase diagram is shown in
the right panel of Fig.\ \ref{fig:phase}. Critical and
multicritical phenomena could be observed in these cases under
external injection of unequal pulses.

The continuation of the first order phase transition line (in all
cases) under external injection to power levels below the one
required for spontaneous passive mode locking (self-starting
threshold) has an interesting practical implication. A laser
operating below the threshold pumping level may be mode locked by
external injection. The injection may be removed and, provided
pumping is strong enough, the laser would remain \emph{metastably}
mode locked  for an exponentially long lifetime \cite{steps}.

\end{document}